\documentclass[preprint,aps,showpacs]{revtex4}
\usepackage{graphicx}
\usepackage{dcolumn}
\usepackage{bm}
\begin{document}
\title{Neutral Higgs Bosons in the SU(3)$_{L}\otimes$U(1)$_{N}$ model}
\author{J. E. Cieza Montalvo}
\affiliation{Instituto de F\'{\i}sica, Universidade do Estado do Rio de Janeiro, Rua S\~ao Francisco Xavier 524, 20559-900 Rio de Janeiro, RJ, Brazil}
\author{M. D. Tonasse}
\affiliation{Campus de Registro, Universidade Estadual Paulista, Rua Tamekishi Takano 5, 11900-000 Registro, SP, Brazil}
\date{\today}

\begin{abstract}
The SU(3)$_L\otimes$U(1)$_N$ electroweak model predicts new Higgs bosons beyond the one of the standard model. In this work we investigate the signature and production of neutral SU(3)$_L\otimes$U(1)$_N$ Higgs bosons in the $e^-e^+$ 
Next Linear Collider (NLC) and in the CERN Linear Collider (CLIC). 
We compute the branching ratios of two of the SU(3)$_L\otimes$U(1)$_N$ neutral Higgs bosons and study the possibility to detect them and the $Z^\prime$ extra neutral boson of the model.
\end{abstract}

\maketitle

\section{Introduction}    

The Higgs sector still remains one of the most indefinite part of the standard model \cite{wsg}, but it still represents a fundamental rule
by explaining how the particles gain masses by means of a isodublet scalar field, which is responsible for the spontaneous breakdown of the gauge symmetry,
the process by which the spectrum of all particles are generated. This process of mass generation is the so called {\it Higgs mechanism}, which plays a central role in gauge theories.
In this process there remains a single neutral scalar, manifesting itself as the Higgs particle. In the standard model only one SU(2) Higgs doublet is necessary and enough to break the gauge symmetry and to generate the particles masses. However,  the standard model does not predicts the number of scalar multiplets of the theory, for that reason, there are several extensions of the standard model containing neutral and charged Higgs bosons. 
The standard model is not able to predicts the mass of the Higgs boson. However, indirect experimental limits are obtained from precision measurements of the electroweak parameters. These measurements are now realized in radiative correction levels, which have a logarithmic dependence of standard Higgs boson mass. From several experiments the present value for the standard Higgs boson mass is $126^{+73}_{-48}$ GeV \cite{lep}. \par
Since the standard model leaves many questions open, there are several well motivated extensions  of it. For example, if the Grand Unified Theory (GUT) contains the standard model at high energies, then the Higgs bosons associated with GUT symmetry breaking must have masses of order $M_{X} \sim {\cal O} (10^{15})$ GeV. Supersymmetry \cite{supers} provides a solution to this hierarchy problem through the cancellation of the quadratic divergences via the contributions of fermionic and bosonic loops \cite{cancell}. Moreover, the Minimal Supersymmetric extension of the Standard Model (MSSM) can be derived as an effective theory from supersymmetric Grand Unified Theories \cite{sgut}. Another promissory class of models is the one based on the $SU(3)_{C}\otimes  SU(3)_{L} \otimes U(1)_{N}$ (3-3-1 for short) semi-simple symmetry group \cite{PT93}. \par
These models emerge as an alternative solution to the problem of violation of unitarity at high energies in processes such as $e^-e^- \to W^-V^-$, induced by right-handed currents coupled to a vector boson $V^-$. The usual way to circumvent this problem is to give particular values to model parameters in order to cancel the amplitude of the process \cite{PP92}, but in \cite{PP92} was proposed an elegant solution assuming the presence of a doubly charged vector boson. The simplest electroweak gauge model that is able to realize naturaly a double charge gauge boson is the one based on the SU(3)$\otimes$U(1) symmetry \cite{PP92}. As a consequence of the extended gauge symmetry, the model is compelled to accommodate a much richer Higgs sector. \par
The main feature of the 3-3-1 model is that it is able to predicts the correct number of fermions families. This is because, contrary to the standard model, the 3-3-1 model is anomalous in each generation. The anomalies are cancelled only if the number of families is a multiple of three. In addition, if we take into account that the asymptotic freedom condition of the QCD is valid only if the number of generations of quarks is to be less than five, we conclude that the number of generations is three \cite{LS01}.  Another good feature is that the model predicts an upper bound for the Weinberg mixing angle at $\sin^{2} {\theta_W} < 1/4$. Therefore, the evolution of $\theta_W$ to high values leads to an upper bound to the new mass scale between 3 TeV and 4 TeV \cite{JJ97}. \par
In this work we are interested in a version of the 3-3-1 model, whose scalar sector has only three Higgs triplets \cite{PT93}. The text is organized as follow. In Sec. II we give the relevant features of the model. In Sec. III we compute the total cross sections of the processes  and the Sec. IV contains our results and conclusions. 

\section{Basic facts about the 3-3-1 model }


The three Higgs triplets of the model are
\begin{equation}
\eta = \left(\begin{array}{c} \eta^0 \\  \eta_1^- \\  \eta_2^+ \end{array}\right), \quad \rho = \left(\begin{array}{c} \rho^+ \\  \rho^0 \\  \rho^{++}
\end{array}\right), \quad 
\chi = \left(\begin{array}{c} \chi^- \\
\chi^{--} \\ \chi^0 \end{array}\right)
\end{equation}
transforming as $\left({\bf 3}, 0\right)$, $\left({\bf 3}, 1\right)$ and $\left({\bf 3}, -1\right)$, respectively. \par
The neutral scalar fields develop the vacuum expectation values (VEVs) $\langle\eta^0\rangle \equiv v_\eta$, $\langle\rho^0\rangle \equiv v_\rho$ and  $\langle\chi^0\rangle \equiv v_\chi$, with $v_\eta^2 + v_\rho^2 = v_W^2 = (246 \mbox{ GeV})^2$. The pattern of symmetry breaking is   
$\mbox{SU(3)}_L \otimes\mbox{U(1)}_N \stackrel{\langle\chi\rangle}{\longmapsto}\mbox{SU(2)}_L\otimes\mbox{U(1)}_Y\stackrel{\langle\eta, \rho\rangle}{\longmapsto}\mbox{U(1)}_{\rm em}$
and so, we can expect $v_\chi \gg v_\eta, v_\rho$. The $\eta$ and $\rho$ scalar triplets give masses to the ordinary fermions and gauge bosons, while the $\chi$ scalar triplet gives masses to the new fermions and new gauge bosons. The most general, gauge invariant and renormalizable Higgs potential is 
\begin{eqnarray}
V\left(\eta, \rho, \chi\right) & = & \mu_1^2\eta^\dagger\eta + \mu_2^2\rho^\dagger\rho + \mu_3^2\chi^\dagger\chi + \lambda_1\left(\eta^\dagger\eta\right)^2 + \lambda_2\left(\rho^\dagger\rho\right)^2 + \lambda_3\left(\chi^\dagger\chi\right)^2 + \cr
&& + \left(\eta^\dagger\eta\right)\left[\lambda_4\left(\rho^\dagger\rho\right) + \lambda_5\left(\chi^\dagger\chi\right)\right] + \lambda_6\left(\rho^\dagger\rho\right)\left(\chi^\dagger\chi\right) + \lambda_7\left(\rho^\dagger\eta\right)\left(\eta^\dagger\rho\right) + \cr
&& + \lambda_8\left(\chi^\dagger\eta\right)\left(\eta^\dagger\chi\right) + \lambda_9\left(\rho^\dagger\chi\right)\left(\chi^\dagger\rho\right) + \lambda_{10}\left(\eta^\dagger\rho\right)\left(\eta^\dagger\chi\right) + \cr
&& \frac{1}{2}\left(f\epsilon^{ijk}\eta_i\rho_j\chi_k + {\mbox{H. c.}}\right).
\label{pot}\end{eqnarray}
Here $\mu_i$ $\left(i = 1, 2, 3\right)$, $f$ are constants with dimension of mass and the $\lambda_i$, $\left(i = 1, \dots, 10\right)$ are dimensionalless constants. $f$ and $\lambda_3$ are negative from the positivity of the scalar masses. The term proportional to $\lambda_{10}$ violates lepto-barionic number,
therefore it was not considered in the analysis of the Ref. \cite{TO96} (another analysis of the 3-3-1 scalar sector are given in Ref. \cite{AK00} and references cited therein). We can notice that this term contributes to the mass matrices of the charged scalar fields, but not to the neutral ones.  However, it can be checked that in the approximation $v_\chi \gg v_\eta, v_\rho$ we can still work with the masses and eigenstates given in Ref. \cite{TO96}. Here this term is important to the decay of the lightest exotic fermion. Therefore, we will keep it in the Higgs potential (\ref{pot}). \par
As usual, symmetry breaking is implemented by shifting the scalar neutral fieldas $\varphi = v_\varphi + \xi_\varphi + i\zeta_\varphi$, with $\varphi$ $=$  $\eta^0$, $\rho^0$, $\chi^0$. Thus, the physical neutral scalar eigenstates  $H^0_1$, $H^0_2$, $H^0_3$ and $h^0$ are related to the shifted fields as    
\begin{equation}  
\left(\begin{array}{c} \xi_\eta \\  \xi_\rho \end{array}\right) \approx
\frac{1}{v_W}\left(\begin{array}{cc} v_\eta & v_\rho \\  v_\rho & -v_\eta 
\end{array}\right)\left(\begin{array}{c} H^0_1 \\  H^0_2 \end{array}\right), \qquad
\xi_\chi \approx H^0_3, \qquad \zeta_\chi  \approx h^0,
\label{eign}
\end{equation}
and in the charged scalar sector we have
\begin{eqnarray}
\eta^+_1 = \frac{v_\rho}{v_W}H^+_1, \quad \rho^+ = \frac{v_\eta}{v_W}H_1^+, \quad \eta^+_2 = \frac{v_\chi}{\sqrt{v_\eta^2 + v_\chi^2}}H_2^+, \quad \chi^+ = \frac{v_\eta}{\sqrt{v_\eta^2 + v_\chi^2}}H^+_2, && \\
\rho^{++} = \frac{v_\chi}{\sqrt{v_\rho^2 + v_\chi^2}}H^{++}, \quad \chi^{++} = \frac{v_\rho}{v_\chi}H^{++}, &&
\label{eigc}
\end{eqnarray}\label{eig}
with the condition that $v_\chi \gg v_\eta, v_\rho$ \cite{TO96}. \par
The content of matter fields form the three SU(3)$_L$ triplets
\begin{equation}
\psi_{aL} = \left(\begin{array}{c} \nu^\prime_{\ell a} \\ \ell^\prime_a \\ P^\prime_a  \end{array}\right), \quad Q_{1L} = \left(\begin{array}{c} u^\prime_1 \\ d^\prime_1 \\ J_1  \end{array}\right), \quad Q_{\alpha L} = \left(\begin{array}{c} J^\prime_\alpha \\ u^\prime_\alpha \\ d^\prime_\alpha  \end{array}\right),
\end{equation}\label{fer}
transform as $\left({\bf 3}, 0\right)$, $\left({\bf 3}, 2/3\right)$ and $\left({\bf 3}^*, -1/3\right)$, respectively, where $\alpha = 2, 3$. In Eqs. (\ref{fer}) $P_a$ are heavy leptons, $\ell^\prime_a = e^\prime, \mu^\prime, \tau^\prime$. The model also predicts the exotic $J_1$ quark, which carries $5/3$ units of elementary electric charge and $J_2$ and $J_3$ with $-4/3$ each. The numbers $0$, $2/3$ and $-1/3$ in Eqs. (\ref{fer}) are the U$_N$ charges. We also have  
the right-handed counterpart of the left-handed matter fields, $\ell^\prime_R \sim \left({\bf 1}, -1\right)$, $P^\prime_R \sim \left({\bf 1}, 1\right)$, $U^\prime_R \sim \left({\bf 1}, 2/3\right)$, $D^\prime_R \sim \left({\bf 1}, -1/3\right)$, $J^\prime_{1R} \sim \left({\bf 1}, 5/3\right)$ and $J^\prime_{2,3R} \sim \left({\bf 1}, -4/3\right)$, where $U = u, c, t$ and $D = d, s, b$ for the ordinary quarks. \par
The Yukawa Lagrangians that respect the gauge symmetry are
\begin{eqnarray}
{\cal L}^Y_\ell & = & -G_{ab}\overline{\psi_{aL}}\ell^\prime_{bR} - G^\prime_{ab}\overline{\psi^\prime_{aL}}P^\prime\chi + {\mbox{H. c.}}, \\
{\cal L}^Y_q & = & \sum_a\left[\overline{Q_1{L}}\left(G_{1a}U^\prime_{aR}\eta + \tilde{G}_{1a}D^\prime_{aR}\rho\right) + \sum_\alpha\overline{Q_{\alpha L}}\left(F_{\alpha a}U^\prime_{aR}\rho^* + \tilde{F}_{\alpha a}D^\prime_{aR}\eta^*\right)\right] + \cr
&& + \sum_{\alpha\beta}F^J_{\alpha\beta}\overline{Q_{\alpha J}}J^\prime_{\beta R}\chi^* + G^J\overline{Q_{1L}}J_{1R} + {\mbox{ H. c.}}
\label{yuk}
\end{eqnarray}
Here, the $G$'s, $\tilde{G}$'s, $F$'s and $\tilde{F}$'s are Yukawa coupling constants with $a, b = 1, 2, 3$ and $\alpha = 2, 3$. \par
It should be noticed that the ordinary quarks couple only through $H^0_1$ and $H^0_2$. This is because these physical scalar states are linear combinations of the interactions eigenstates $\eta$ and $\rho$, which break the SU(2)$_L$$\otimes$U(1)$_Y$ symmetry  to U(1)$_{\rm em}$. On the other hand the heavy-leptons and quarks couple only through $H^0_3$ and $h^0$ in scalar sector, {\it i. e.}, throught the Higgs that induces the symmetry breaking of SU(3)­$_L$$\otimes$U(1)$_N$ to SU(2)$_L$$\otimes$U(1)$_Y$. The Higgs particle spectrum consists of seven states: three scalars $\left(H_{1}^{0}, H_{2}^{0}, H_3^{0}\right)$, one neutral pseudoscalar $h^0$ and three charged Higgs bosons, $H_{1}^{+}$, $H_{2}^{+}$ and $H^{++}$. \par 
In this work we study the production of a neutral Higgs boson at $e^{-} e^{+}$ colliders because of lower backgrounds and since it is one of the most promising in the search for the Higgs. The Higgs $H_{i}$, where $i = 1, 2$ can be radiated from a $Z$ and $Z^\prime$ boson. The $Z (Z^\prime) \to Z  \ H_{1} (H_{2})$ process is the dominant mechanism at the $Z$ resonance energy. We discuss this process only for on-shell $Z$ production. In this work, we will study the production mechanism for Higgs particles in $e^{+} e^{-}$ colliders such as the Next Linear Collider (NLC) ($\sqrt{s} = 500$ GeV) and CERN Linear Collider (CLIC) ($\sqrt{s} = 1000$ GeV). 

\section{Cross section production }       


The main mechanism for the production of Higgs particles in $e^+e^-$ collisions occurs in association with the boson $Z$, and $Z^\prime$ through the Drell-Yan mechanism. The process $e^{+} e^{-} \to  H_{i} Z$ $(i = 1, 2)$ takes place through the exchange of bosons $Z$ and $Z^\prime$ in the $s$ channel. Then using the interaction Lagrangian (\ref{yuk}) and (\ref{pot}) we obtain the differential cross section 
\begin{eqnarray} 
\frac{d \hat{\sigma}}{d\cos \theta} & = & \frac{\beta_H \alpha^{2} \pi}{32 \sin^{4}_{\theta_{W}} \cos^{2}_{\theta_{W}} s}\times \cr
&& \times\left[\frac{cZH_i^{2}}{\left(s- M_{Z}^{2} + i M_Z 
\Gamma_Z\right)^{2}}\left(2M_{Z}^{2}+ \frac{2tu}{M_{Z}^{2}}- 2t- 2u + 2s\right)\left(g_{V}^{e^{2}}+ g_{A}^{e^{2}}\right) + \nonumber  \right.\\ 
&& \left. + \frac{cZ^\prime H_i^{2}}{\left(s- M_{Z\prime}^{2}+ iM_{Z^\prime} \Gamma_{Z^\prime}\right)^{2}}\left(2M_{Z}^{2}+ \frac{2tu}{M_{Z}^{2}}- 2t- 2u + 2s\right)\left(g_{V^\prime}^{e^{2}}+ g_{A^\prime}^{e^{2}}\right) +  \nonumber \right. \\
&& \left. + \frac{2 \ cZH_icZ^\prime H_i}{\left(s- M_{Z}^{2}+ iM_{Z} \Gamma_{Z}\right)\left(s- M_{Z^\prime}^{2}+ iM_{Z^\prime} \Gamma_{Z^\prime}\right)}\times \right. \cr
&& \left. \times\left(2M_{Z}^{2}+ \frac{2tu}{M_{Z}^{2}}- 2t- 2u + 2s\right)\left(g_{V}^{e}g_{V^\prime}^{e}+ g_{A}^{e} g_{A^\prime}^{e}\right)\right].
\label{DZZH}
\end{eqnarray} 
The primes $\left(^\prime\right)$ are for the case when we take a $Z^\prime$ boson, $\Gamma_{Z}$ and $\Gamma_{Z^\prime}$ \cite{cieto}, are the total width of the $Z$ and $Z^\prime$ boson, $g_{V, A}^{e}$ are the standard lepton coupling constants, $g_{V^\prime, A^\prime}^{e}$ are the $3-3-1$ lepton coupling constants, $\sqrt{s}$ is the center of mass energy of the $e^{-} e^{+}$ system. For the $Z^\prime$ boson we take  $M_{Z^\prime} = \left(0.5 - 3\right)$ TeV, since $M_{Z^\prime}$ is proportional to the VEV $v_\chi$ \cite{PP92,FR92}. For the standard model parameters we assume PDG values, {\it i. e.}, $M_Z = 91.19$ GeV, $\sin^2{\theta_W} = 0.2315$, and $M_W = 80.33$ GeV 
\cite{Cea98}, the velocity of the Higgs in the center of mmass (CM) of the  process is denoted through  $\beta_{H}$, t and u are the kinemetic invariants, the $cZZH_{i}^{0} (cZZ^\prime H_{i}^{0})$ are the coupling constants of the $Z$ boson to $Z(Z^\prime)$ bosons and Higgs $H_{i}^{0}$ where i stands for $H_{1}^{0}, H_{2}^{0}$, the cHi0VPM are the coupling constants of the $H_{i}^{0}$, where i= 1,2, to $V^{-}V^{+}$, the cHi0UPP are the coupling constants of the $H_{i}^{0}$, where i=1,2, to $U^{--}U^{++}$, and the cHi0HPP are the coupling constants of the $H_{i}^{0}$, where i=1,2, to $H^{--}H^{++}$. We then have that  
\begin{eqnarray}
t & = & M_{Z}^{2} - \frac{s}{2}\left\{1 + \frac{M_{Z}^{2}- M_{H}^{2}}{s} + \right. \\ && \left. -\cos \theta  \left[\left( 1- \frac{\left(M_{Z}+ M_{H}\right)^{2}}{s} \right) \left(1- \frac{\left(M_{Z}- M_{H}\right)^{2}}{s} \right) \right ]^{1/2}\right\}, \\
u & = & M_{H}^{2} - \frac{s}{2}\left\{1 - \frac{M_{Z}^{2}- M_{H}^{2}}{s} + \right. \\ && \left. + \cos \theta  \left [\left( 1- \frac{\left(M_{Z}+ M_{H}\right)^{2}}{s} \right) \left(1- \frac{\left(M_{Z}- M_{H}\right)^{2}}{s} \right) \right ]^{1/2}\right\}, \\
cZZ^\prime H_{1}^{0} & = & -i\frac{g^{2}}{2\sqrt{3} v_{W}} \frac{M_{Z}}{M_{W}} \ \frac{\left[v_{\eta}^{2} \left(6 t_{W}^{2}+1\right)- v_{\rho}^{2}\right]}{\sqrt{1+ 3t_{W}^{2}}}, \\
cZZ^\prime H_{2}^{0} & = & i\frac{g^{2}}{\sqrt{3}} \ \frac{v_{\eta} v_{\rho}}{v_{W}} \sqrt{1+ 4t_{W}^{2}}, \quad cH10VPM = -i g^{2}\frac{v^{2}_{\rho}}{v_{W}}, \\ 
cH20VPM & = & -i g^{2} \frac{v_{\eta} v_{\rho}}{v_{W}}, \quad cH10UPP = -i g^{2} \frac{v^{2}_{\eta}}{v_{W}}, \quad cH20UPP = i g^{2} \frac{v_{\eta} v_{\rho}}{v_{W}}, \\
cH10HPP & = & -i\frac{2\left[\left(\lambda_{6}+ \lambda_{9}\right) v_{\eta}^{4}+ \left(2 \lambda_{2}+ \lambda_{9}\right)v_{\eta}^{2} v_{\chi}^{2}+ \left(\lambda_{4}+ \lambda_{5}\right) v_{\eta}^{2} v_{\chi}^{2}\right]- f v_{\eta} v_{\rho} v_{\chi}}{v_{W}\left(v_{\eta}^{2}+ v_{\chi}^{2}\right)}, \\
cH20PP & = & iv_{\eta} \frac{2 v_{\rho} \left[\left(2\lambda_{2}-  \lambda_{4}+ \lambda_{9}\right) v_{\chi}^{2} + \left(\lambda_{6}- \lambda_{5}+ \lambda_{9}\right) v_{\eta}^{2} \right]- f v_{\eta} v_{\chi}}{v_{W}\left(v_{\eta}^{2}+ v_{\chi}^{2}\right)},
\end{eqnarray}  
where $\theta$ is the angle between the Higgs and the incident electron  in the CM frame, where the coupling constant of the $Z$ boson to $Z$ and $H_{1}^{0}$ are the standard ones and the coupling constant of the $Z$ boson to $Z$ and $H_{2}^{0}$ does not exist. \par
The total width of the Higgs $H_{1}^{0}$ into quarks, leptons, $W^{+}W^{-}$, $ZZ$, $ZZ^\prime$, $Z^\prime Z^\prime$ gauge bosons, $H_{2}^{0} \ H_{2}^{0}$, $H_{1}^{-} H_{1}^{+}$, $H_{2}^{-} H_{2}^{+}$, $h^{0}   h^{0}$, $H_{2}^{0}  H_{3}^{0}$ Higgs bosons, $V^{-}V^{+}$ charged bosons, $U^{--} U^{++}$ double charged bosons, $H_{2}^{0} Z$, $H_{2}^{0}Z^\prime$ bosons and $H^{--} H^{++}$ double charged Higgs bosons, are,
respectively, given by
\begin{eqnarray}
\Gamma \left(H_{1}^{0} \to {\rm all}\right)&  = & \Gamma_{H_{1}^{0} \to q \bar{q}}  +   \Gamma_{H_{1}^{0} \to \ell^{-} \ell^{+}} +  \Gamma_{H_{1}^{0} \to W^{+} W^{-}} + \Gamma_{H_{1}^{0} \to Z Z} + \Gamma_{H_{1}^{0} \to Z^\prime Z}+  \Gamma_{H_{1}^{0} \to Z^\prime Z^\prime} + \nonumber  \\
&&+ \Gamma_{H_{1}^{0} \to H^{0}_{2} H^{0}_{2}} + \Gamma_{H_{1}^{0} \to H^{-}_{1} H^{+}_{1}} + \Gamma_{H_{1}^{0} \to H^{-}_{2} H^{+}_{2}} + \Gamma_{H_{1}^{0} \to h^{0} h^{0}} + \Gamma_{H_{1}^{0} \to H^{0}_{2}  H^{0}_{3}} +      \nonumber \\
&& \Gamma_{H_{1}^{0} \to V^{-} V^{+}} + \Gamma_{H_{1}^{0} \to U^{-} U^{+}} +  \Gamma_{H_{1}^{0} \to H_{2}^{0} Z} + \Gamma_{H_{1}^{0} \to H_{2}^{0} Z^\prime} + \Gamma_{H_{1}^{0} \to H^{--} H^{++}}, 
\end{eqnarray}
where we have for each the widths given above that
\begin{eqnarray}  
\Gamma_{H_{1}^{0} \to q \bar{q}} & = & \frac{3 \sqrt{1- 4 M_{q}^2/  M^{2}_{H_{1}^{0}}}}{16 \pi M_{H_{1}^{0}}} \frac{M_{q}^{2}}{v_{W}^{2}} (M^{2}_{H_{1}^{0}} - 2 M_{q}^{2} ) ,  \\
\Gamma_{H_{1}^{0} \to \ell{-} \ell{+}} & = & \frac{\sqrt{1- 4 M_{\ell}^2/  M^{2}_{H_{1}^{0}}}}{16 \pi M_{H_{1}^{0}}} \frac{M_{\ell}^{2}}{v_{W}^{2}}  (M^{2}_{H_{1}^{0}} - 2 M_{\ell}^{2} ) ,  \\
\Gamma_{H_{1}^{0} \to W{-} W{+}} & = & \frac{\sqrt{1- 4 M_W^2/  M^{2}_{H_{1}^{0}}}}{8 \pi} \frac{g^{2} M_W^2}{M_{H_{1}^{0}}} \left (3 - \frac{M^{2}_{H_{1}^{0}}}{M_W^2}+ \frac{M^{4}_{H_{1}^{0}}}{4 M_W^4} \right ),  \\
\Gamma_{H_{1}^{0} \to ZZ} & = & \frac{\sqrt{1- 4 M_Z^2/  M^{2}_{H_{1}^{0}}}}{8 \pi \cos^{2}_{\theta_{W}}} \frac{g^{2} M_Z^2}{M_{H_{1}^{0}}} \left (3 - \frac{M^{2}_{H_{1}^{0}}}{M_Z^2}+ \frac{M^{4}_{H_{1}^{0}}}{4 M_Z^4} \right ) ,  \\  
\Gamma_{H_{i}^{0} \to Z^\prime Z} & = & \frac{\sqrt{1- \left ( \frac{M_{Z^\prime}+ M_{Z}}{M_{H_{i}^{0}}} \right )^{2}} \sqrt{1- \left ( \frac{M_{Z^\prime}- M_{Z}}{M_{H_{i}^{0}}} \right )^{2}}}{8 \pi M_{H_{i}^{0}}} (cZZ^\prime H_{i}^{0})^{2} \times \cr
&& \times\left (\frac{5}{2}+ \frac{1}{4} \frac{M_{Z}^{2}}{M_{Z^\prime}^{2}} + \frac{1}{4} \frac{M_{Z^\prime}^{2}}{M_{Z}^{2}} + \frac{1}{4} \frac{M_{{H}_{i}^{0}}^{4}}{M_{Z^\prime}^{2} M_{Z}^{2}} - \frac{1}{2} 
\frac{M_{{H}_{i}^{0}}^{2}}{M_{Z}^{2}} - \frac{1}{2} \frac{M_{{H}_{i}^{0}}^{2}}{M_{Z^\prime}^{2}} \right ) ,   \\
\Gamma_{H_{1}^{0} \to Z^\prime Z^\prime} & = & \frac{\sqrt{1- 4 M_Z^{\prime2}/  M^{2}_{H_{1}^{0}}} \ g^{4} (1+ 3 t_{W}^{2})^{2} (12 t_{W}^{2} v_{\eta}^{2} (1+ 3 t_{W}^{2}) + v_{\chi}^{2})^{2}}{576 \pi v_{\chi}^{2} M_{{H}_{1}^{0}}}\times \cr 
&& \times\left (3- \frac{M_{{H}_{1}^{0}}^{2}}{M_{Z^\prime}^{2}}+ \frac{1}{4}\frac{M_{H_{1}^{0}}^{4}}{M_{Z^\prime}^{4}} \right ),  \\  
\Gamma_{H_{1}^{0} \to H_{2}^{0} H_{2}^{0}} & = & \frac{\sqrt{1- 4 M_{H_{2}^{0}}^2/  M^{2}_{H_{1}^{0}}}}{4 \pi M_{H_{1}^{0}}} \left ( \frac{\lambda_{4}(v_{\eta}^{4} + v_{\rho}^{4})+ 2 v_{\eta}^{2} v_{\rho}^{2} (2 \lambda_{4}+ 2 \lambda_{2}+ 3 \lambda_{1})}{v_{\chi}^{3}} \right )^{2},   \\
\Gamma_{H_{1}^{0} \to H_{1}^{-} H_{1}^{+}} & = & \frac{\sqrt{1- 4 M_{H_{1}^{\pm}}^2/  M^{2}_{H_{1}^{0}}}}{4 \pi M_{H_{1}^{0}}} \left ( \frac{ (\lambda_{4}+ \lambda_{7})(v_{\eta}^{4} + v_{\rho}^{4})+ 2 v_{\eta}^{2} v_{\rho}^{2} (\lambda_{1}+ \lambda_{2}+ \lambda_{7})}{v_{\chi}^{3}} \right )^{2},   \\
\Gamma_{H_{1}^{0} \to H_{2}^{-} H_{2}^{+}} & = & \frac{\sqrt{1- 4 M_{H_{2}^{\pm}}^2/  M^{2}_{H_{1}^{0}}}}{16 \pi M_{H_{1}^{0}}} \times\cr
&& \hskip -1 true cm \times\left ( \frac{ (-\lambda_{4} v_{\chi}^{2} + \lambda_{6} v_{\rho}^{2}) v_{\eta}^{2}- 2(\lambda_{5} + \lambda_{8}) v_{\rho}^{4}- 2(2 \lambda_{1}+ \lambda_{8}) v_{\rho}^{2} v_{\chi}^{2}+ f v_{\eta} v_{\rho} v_{\chi}}{v_{W}  (v_{\rho}^{2}+ v_{\chi}^{2})} \right )^{2},   \\
\Gamma_{H_{1}^{0} \to h_{0} h_{0}} & = & \frac{\sqrt{1- 4 M^{2}_{h_{0}}/  M^{2}_{H_{1}^{0}}}}{4 \pi M_{H_{1}^{0}}} \left (\frac{\lambda_{5} v_{\rho}^{2} + \lambda_{6} v_{\eta}^{2}}{v_{\chi}} \right )^{2},  \\
\Gamma_{H_{i}^{0} \to H_{2}^{0} H_{3}^{0}} & = & \frac{\sqrt{1- \left ( \frac{M_{H_{2}^{0}}+ M_{H_{3}^{0}}}{M_{H_{i}^{0}}} \right )^{2}} \sqrt{1- \left ( \frac{M_{H_{2}^{0}}- M_{H_{3}^{0}}}{M_{H_{i}^{0}}} \right )^{2}}}{16 \pi M_{H_{i}^{0}}}\times \\
&& \times\left(\frac{4 (\lambda_{5}- \lambda_{6}) v_{\eta} v_{\rho} v_{\chi} + f ( v_{\eta}^{2}- v_{\rho}^{2})}{v_{\chi}^{2}} \right )^{2} ,  \\
\Gamma_{H_{i}^{0} \to V^{-} V^{+}} & = & \frac{\sqrt{1- 4 M_{V^{\pm}}^2/  M^{2}_{H_{i}^{0}}}}{8 \pi M_{H_{i}^{0}}} (cHi0VPM)^{2} \left (3 - \frac{M^{2}_{H_{i}^{0}}}{M_{V^{\pm}}^2}+ \frac{M^{4}_{H_{i}^{0}}}{4 M_{V^{\pm}}^4} \right ) ,  \\ 
\Gamma_{H_{i}^{0} \to U^{--} U^{++}} & = & \frac{\sqrt{1- 4 M_{U^{\pm \pm}}^2/  M^{2}_{H_{i}^{0}}}}{8 \pi M_{H_{i}^{0}}} (cHi0UPP)^{2} \left (3 - \frac{M^{2}_{H_{i}^{0}}}{M_{U^{\pm \pm}}^2}+ \frac{M^{4}_{H_{i}^{0}}}{4 M_{U^{\pm \pm}}^4} \right ) ,  \\  
\Gamma_{H_{1}^{0} \to H_{2}^{0} Z} & = & \frac{\sqrt{1- \left ( \frac{M_{H_{2}^{0}} +  M_{Z}}{M_{H_{1}^{0}}} \right )^{2}} \sqrt{1- \left ( \frac{M_{H_{2}^{0}}- M_{Z}}{M_{H_{1}^{0}}} \right )^{2}}}{4 \pi M_{H_{1}^{0}}} \left (\frac{g M_{Z} v_{\eta} v_{\rho}}{M_{W} v_{\chi}} \right )^{2} \times \cr
&& \times  \left (\frac{M_{Z}^{2}}{4} + \frac{M_{H_{2}^{0}}^{2}}{4M_{Z}^{2}}- \frac{M_{H_{2}^{0}}^{2} M_{H_{1}^{0}}^{2}}{2M_{Z}^{2}} + \frac{M_{H_{1}^{0}}^{4}}{4M_{Z}^{2}} - \frac{M_{H_{2}^{0}}^{2}}{2} - \frac{M_{H_{1}^{0}}^{2}}{2} \right ) ,  \\ 
\Gamma_{H_{1}^{0} \to H_{2}^{0} Z^\prime} & = & \frac{3\sqrt{1- \left ( \frac{M_{H_{2}^{0}}+ M_{Z^\prime}}{M_{H_{1}^{0}}} \right )^{2}} \sqrt{1- \left ( \frac{M_{H_{2}^{0}}- M_{Z^\prime}}{M_{H_{1}^{0}}} \right )^{2}}}{4 \pi M_{H_{1}^{0}}} \left (\frac{g v_{\eta} v_{\rho}t_{W}^{2}}{v_{\chi}^{2} \sqrt{1+ 3 t_{W}^{2}}} \right )^{2} \times \cr
&& \times \left (\frac{M_{Z^\prime}^{2}}{4} + \frac{M_{H_{2}^{0}}^{2}}{4M_{Z^\prime}^{2}} - \frac{M_{H_{2}^{0}}^{2} M_{H_{1}^{0}}^{2}}{2M_{Z^\prime}^{2}} + \frac{M_{H_{1}^{0}}^{4}}{4M_{Z^\prime}^{2}}- \frac{M_{H_{2}^{0}}^{2}}{2} -\frac{M_{H_{1}^{0}}^{2}}{2} \right ) ,  \\ 
\Gamma_{H_{i}^{0} \to H^{--} H^{++}} & = & \frac{\sqrt{1- 4 M^{2}_{H^{\pm \pm}}/  M^{2}_{H_{i}^{0}}}}{16 \pi M_{H_{i}^{0}}} (cHi0HPP)^{2} ,
\end{eqnarray}
where using (15) to (22) and putting i=1,2, we will have the total width for $H_{1}^{0}$ and part of the total width for $H_{2}^{0}$. The total width of the Higgs $H_{2}^{0}$ into quarks, leptons, Z Z$^{\prime}$, Z$^{\prime}$Z$^{\prime}$ gauge bosons,  $H_{1}^{-} H_{1}^{+}$, $H_{2}^{-} H_{2}^{+}$, $h^{0}   h^{0}$, $H_{1}^{0}  H_{3}^{0}$ higgs bosons, $V^{-}V^{+}$ charged bosons, $U^{--} U^{++}$ double charged bosons, $H_{1}^{0} Z$, $H_{1}^{0} Z^\prime$ bosons and $H^{--} H^{++}$ double charged Higgs bosons, is given by
\begin{eqnarray}
\Gamma \left(H_{2}^{0} \to {\rm all}\right)&  = & \Gamma_{H_{2}^{0} \to q \bar{q}}  +   \Gamma_{H_{2}^{0} \to \ell^{-} \ell^{+}} + \Gamma_{H_{2}^{0} \to Z^\prime Z}+  \Gamma_{H_{2}^{0} \to Z^\prime Z^\prime}+ \Gamma_{H_{2}^{0} \to H^{-}_{1} H^{+}_{1}} + \cr
&& + \Gamma_{H_{2}^{0} \to H^{-}_{2} H^{+}_{2}} + \Gamma_{H_{2}^{0} \to h^{0} h^{0}} + \Gamma_{H_{2}^{0} \to H^{0}_{1}  H^{0}_{3}} + \Gamma_{H_{2}^{0} \to V^{-} V^{+}} + \Gamma_{H_{2}^{0} \to U^{-} U^{+}} + \cr
&& +  \Gamma_{H_{2}^{0} \to H_{1}^{0} Z} + \Gamma_{H_{2}^{0} \to H_{1}^{0} Z^\prime} + \Gamma_{H_{2}^{0} \to H^{--} H^{++} }, 
\end{eqnarray}
where we have for the remaining part of the $H_{2}^{0}$
\begin{eqnarray}  
\Gamma_{H_{2}^{0} \to b \bar{b}} & = & \frac{3 \sqrt{1- 4 M_{b}^2/  M^{2}_{H_{2}^{0}}}}{16 \pi M_{H_{2}^{0}}} \frac{M_{b}^{2} v_{\rho}^{2}}{v_{W}^{2} v_{\eta}^{2}} (M^{2}_{H_{2}^{0}} - 2 M_{b}^{2} ) ,  \\
\Gamma_{H_{2}^{0} \to c \bar{c}(t \bar{t})} & = & \frac{3 \sqrt{1- 4 M_{c}^2/  M^{2}_{H_{2}^{0}}}}{16 \pi M_{H_{2}^{0}}} \frac{M_{c}^{2} v_{\eta}^{2}}{v_{W}^{2} v_{\rho}^{2}} (M^{2}_{H_{2}^{0}} - 2 M_{c}^{2} ) ,  \\
\Gamma_{H_{2}^{0} \to \tau^{-} \tau^{+}} & = & \frac{\sqrt{1- 4 M_{\tau}^2/  M^{2}_{H_{2}^{0}}}}{16 \pi M_{H_{2}^{0}}} \frac{M_{\tau}^{2} v_{\rho}^{2}}{v_{W}^{2} v_{\eta}^{2}}  (M^{2}_{H_{1}^{0}} - 2 M_{\tau}^{2} ) ,  \\
\Gamma_{H_{2}^{0} \to Z^\prime Z^\prime} & = & \frac{\sqrt{1- 4 M_Z^{\prime2}/  M^{2}_{H_{2}^{0}}} \ g^{4}  t_{W}^{4}}{\pi M_{{H}_{2}^{0}}}  \frac{v_{\eta}^{2} v_{\rho}^{2}}{v_{\chi}^{2}}\left(3- \frac{M_{{H}_{2}^{0}}^{2}}{M_{Z^\prime}^{2}}+ \frac{1}{4}\frac{M_{H_{2}^{0}}^{4}}{M_{Z^\prime}^{4}} \right ),  \\  
\Gamma_{H_{2}^{0} \to H_{1}^{-} H_{1}^{+}} & = & \frac{\sqrt{1- 4 M_{H_{1}^{\pm}}^2/  M^{2}_{H_{2}^{0}}}}{4 \pi M_{H_{2}^{0}}}\left(v_{\eta} v_{\rho} \frac{ (\lambda_{4}- 2\lambda_{1})v_{\eta}^{2}+ v_{\rho}^{2} (2\lambda_{2}- \lambda_{4})}{v_{\chi}^{3}} \right )^{2},   \\
\Gamma_{H_{2}^{0} \to H_{2}^{-} H_{2}^{+}} & = & \frac{\sqrt{1- 4 M_{H_{2}^{\pm}}^2/  M^{2}_{H_{2}^{0}}}}{16 \pi M_{H_{2}^{0}}}\times \cr
&& \times \left (v_{\rho} \frac{2 (\lambda_{5} + \lambda_{8} - \lambda_{6}) v_{\eta} v_{\rho}^{2}+ 2(2\lambda_{1} - \lambda_{4}) v_{\eta} v_{\chi}^{2}+ 2 \lambda_{8} v_{\rho} v_{\chi}^{2}+ f v_{\rho} v_{\chi}}{v_{W} (v_{\rho}^{2}+ v_{\chi}^{2})} \right )^{2},   \\
\Gamma_{H_{2}^{0} \to h_{0} h_{0}} & = & \frac{\sqrt{1- 4 M^{2}_{h_{0}}/  M^{2}_{H_{2}^{0}}}}{4 \pi M_{H_{2}^{0}}} \left (\frac{(\lambda_{6}- \lambda_{5}) v_{\eta} v_{\rho}}{v_{\chi}} \right )^{2},  \\
\Gamma_{H_{2}^{0} \to H_{1}^{0} Z} & = & \frac{\sqrt{1- \left ( \frac{M_{H_{1}^{0}} +  M_{Z}}{M_{H_{2}^{0}}} \right )^{2}} \sqrt{1- \left ( \frac{M_{H_{1}^{0}}- M_{Z}}{M_{H_{2}^{0}}} \right )^{2}}}{4 \pi M_{H_{2}^{0}}} \left (\frac{g M_{Z} v_{\eta} v_{\rho}}{M_{W} v_{\chi}} \right )^{2} \times \cr
&& \times  \left (\frac{M_{Z}^{2}}{4} + \frac{M_{H_{1}^{0}}^{2}}{4M_{Z}^{2}} - \frac{M_{H_{1}^{0}}^{2} M_{H_{2}^{0}}^{2}}{2M_{Z}^{2}} + \frac{M_{H_{2}^{0}}^{4}}{4M_{Z}^{2}}- \frac{M_{H_{1}^{0}}^{2}}{2} - \frac{M_{H_{2}^{0}}^{2}}{2} \right ) ,  \\ 
\Gamma_{H_{2}^{0} \to H_{1}^{0} Z^\prime} & = & \frac{3 \sqrt{1- \left ( \frac{M_{H_{1}^{0}}+ M_{Z^\prime}}{M_{H_{2}^{0}}} \right )^{2}} \sqrt{1- \left ( \frac{M_{H_{1}^{0}}- M_{Z^\prime}}{M_{H_{2}^{0}}} \right )^{2}}}{4 \pi M_{H_{2}^{0}}} \left (\frac{g v_{\eta} v_{\rho}t_{W}^{2}}{v_{\chi}^{2} \sqrt{1+ 3 t_{W}^{2}}} \right )^{2} \times \cr
&& \times \left (\frac{M_{Z^\prime}^{2}}{4} + \frac{M_{H_{2}^{0}}^{2}}{4M_{Z^\prime}^{2}} - \frac{M_{H_{2}^{0}}^{2} M_{H_{1}^{0}}^{2}}{2M_{Z^\prime}^{2}} + \frac{M_{H_{1}^{0}}^{4}}{4M_{Z^\prime}^{2}}- \frac{M_{H_{2}^{0}}^{2}}{2} - \frac{M_{H_{1}^{0}}^{2}}{2} \right ) ,  
\end{eqnarray}
The total width of the $Z^\prime$ boson, whose one part was already calculated in \cite{cieto}, is
\begin{eqnarray}  
\Gamma \left(Z^\prime \to {\rm all}\right)& = & \Gamma_{Z^\prime \to P^{-}  P^{+}} +  
\Gamma_{Z^\prime \to \ell^{-}_{i} \ell^{+}_{i}} + \Gamma_{Z^\prime \to \nu_{i}  \bar{\nu}_{i}} + \Gamma_{Z^\prime \to q \bar{q} \left(J \bar{J}\right)} +  \Gamma_{Z^\prime \to X^{-} X^{+}} + \Gamma_{Z^\prime \to H_{1}^{0} H_{1}^{0}} + \nonumber   \\
&& + \Gamma_{Z^\prime \to H_{2}^{0} H_{2}^{0}}+ \Gamma_{Z^\prime \to H_{1}^{-} H_{1}^{+}} + \Gamma_{Z^\prime \to H_{2}^{-} H_{2}^{+}}+  \Gamma_{Z^\prime \to H_{1}^{0} H_{2}^{0}}+  \Gamma_{Z^\prime \to H_{1}^{0} Z} + \Gamma_{Z^\prime \to H_{2}^{0} Z}  \nonumber   ,  
\end{eqnarray}
where $i=e, \mu$ and $\tau$, $X^\pm = V^\pm$ or $U^{\pm\pm}$ and we have for the others particles the relations 
\begin{eqnarray}    
\Gamma_{Z^\prime \to H_{1}^{0} H_{1}^{0}} & = & \frac{\sqrt{1- 4 M^{2}_{H_{1}^{0}}/  M^{2}_{Z^\prime}}}{2304 \pi M_{Z^\prime}} \left (g \frac{v_{W}^{2}+ 6 v_{\eta}^{2} t_{W}^{2}}{v_{W}^{2}(1+ 3t_{W}^{2})} \right )^{2} (M_{Z^\prime}^{2}- 4 M^{2}_{H_{1}^{0}}),  \\
\Gamma_{Z^\prime \to H_{2}^{0} H_{2}^{0}} & = & \frac{\sqrt{1- 4 M^{2}_{H_{2}^{0}}/  M^{2}_{Z^\prime}}}{2304 \pi M_{Z^\prime}} \left (g \frac{v_{W}^{2}+ 6 v_{\eta}^{2} t_{W}^{2}}{v_{W}^{2}(1+ 3t_{W}^{2})} \right )^{2} (M_{Z^\prime}^{2}- 4 M^{2}_{H_{2}^{0}}) ,  \\
\Gamma_{Z^\prime \to H_{1}^{-} H_{1}^{+}} & = & \frac{\sqrt{1- 4 M^{2}_{H_{1}^{\pm}}/  M^{2}_{Z^\prime}}}{576 \pi M_{Z^\prime}} \left (\frac{g v_{\rho}^{2} (1+ 6 t_{W}^{2}) + v_{\eta}^{2}}{v_{\chi}^{2}(1+ 3 t_{W}^{2})} \right )^{2} (M_{Z^\prime}^{2}- 4 M^{2}_{H_{1}^{\pm}}) ,  \\
\Gamma_{Z^\prime \to H_{2}^{-} H_{2}^{+}} & = & \frac{\sqrt{1- 4 M^{2}_{H_{2}^{\pm}}/  M^{2}_{Z^\prime}}}{576 \pi M_{Z^\prime}} \left (\frac{g v_{\rho}^{2} (1+ 6 t_{W}^{2}) + v_{\eta}^{2}}{v_{\chi}^{2}(1+ 3 t_{W}^{2})} \right )^{2} (M_{Z^\prime}^{2}- 4 M^{2}_{H_{2}^{\pm}}).
\end{eqnarray}

\section{Results and conclusions }    


In the following we present the cross section for the process $e^{-} e^{+} \to  Z  H^{0}_{i}$ where $i=1,2$, for the NLC (500 GeV) and CLIC (1000 GeV). In all calculations it will be taken the following parameters $M_{J_{1}} = 250$ GeV, $M_{J_{2}} = 350$ GeV, $M_{J_{3}} = 500$ GeV, $M_{V}^{\pm} = 200$ GeV, $M_{U^{\pm \pm}}=200$ GeV, $M_{P_{a}}=200$ GeV, $M_{Z^\prime} = 600$ GeV, $\lambda_{i}=1$ where $i=1,2,...,9$, $M_{H_{i}^{0}}=200$ GeV where $i=1,2,3$, $M_{H_{i}^{\pm}}=200$ GeV where $i=1,2$, $M_{H^{++}}=200$ GeV, $f=-1000$ GeV and the vacuum expectation value $w=1000$ GeV. 
\begin{figure}
\scalebox{0.7}{\includegraphics*[0cm, 2.9cm][20cm,12cm]{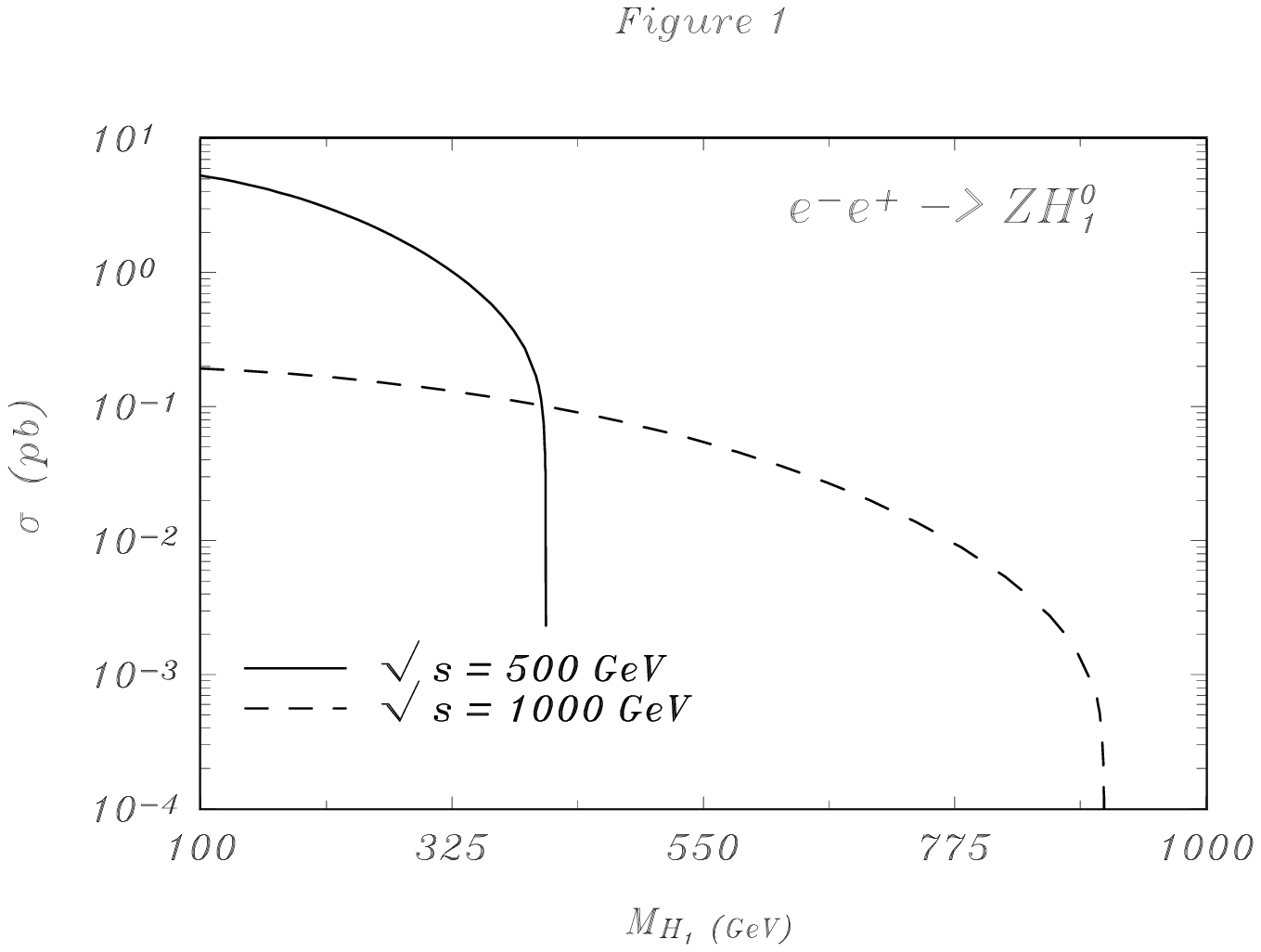}}
\caption{\label{fig1} \footnotesize Total cross section for the process $e^{-} e^{+} \rightarrow ZH_{1}^{0}$ as a 
function of $M_{H_{1}^{0}}$ for a $v_{\eta}=140$ GeV at $\sqrt{s} = 500$ GeV (solid line) and $\sqrt{s} = 1000$ GeV (dashed line).}
\end{figure}
The mass of $M_{Z^{\prime}}$ taken above is in accord with the estimates of the CDF and D0 experiments, which probes the $Z^{\prime}$ masses in the 500-800 GeV range, \cite{tait}, while the reach of the LHC is superior for higher masses, that is $1 TeV <M_{Z^{\prime}} \leq 5$ TeV, \cite{freitas}. With regards to Higgs the LHC is able to discover the Higgs boson with a mass up to 1 TeV and to check its basic properties. In Fig. 1, we show the cross section $e^{-} e^{+} \to  Z H_{1}^{0}$, this process will be studied in two cases, the one where we put for the vacuum expectation value $v_{\eta}=140$ GeV and the other $v_\eta = 240$ GeV, respectively. \begin{figure}
\scalebox{0.7}{\includegraphics*[0cm, 2.9cm][20cm,12cm]{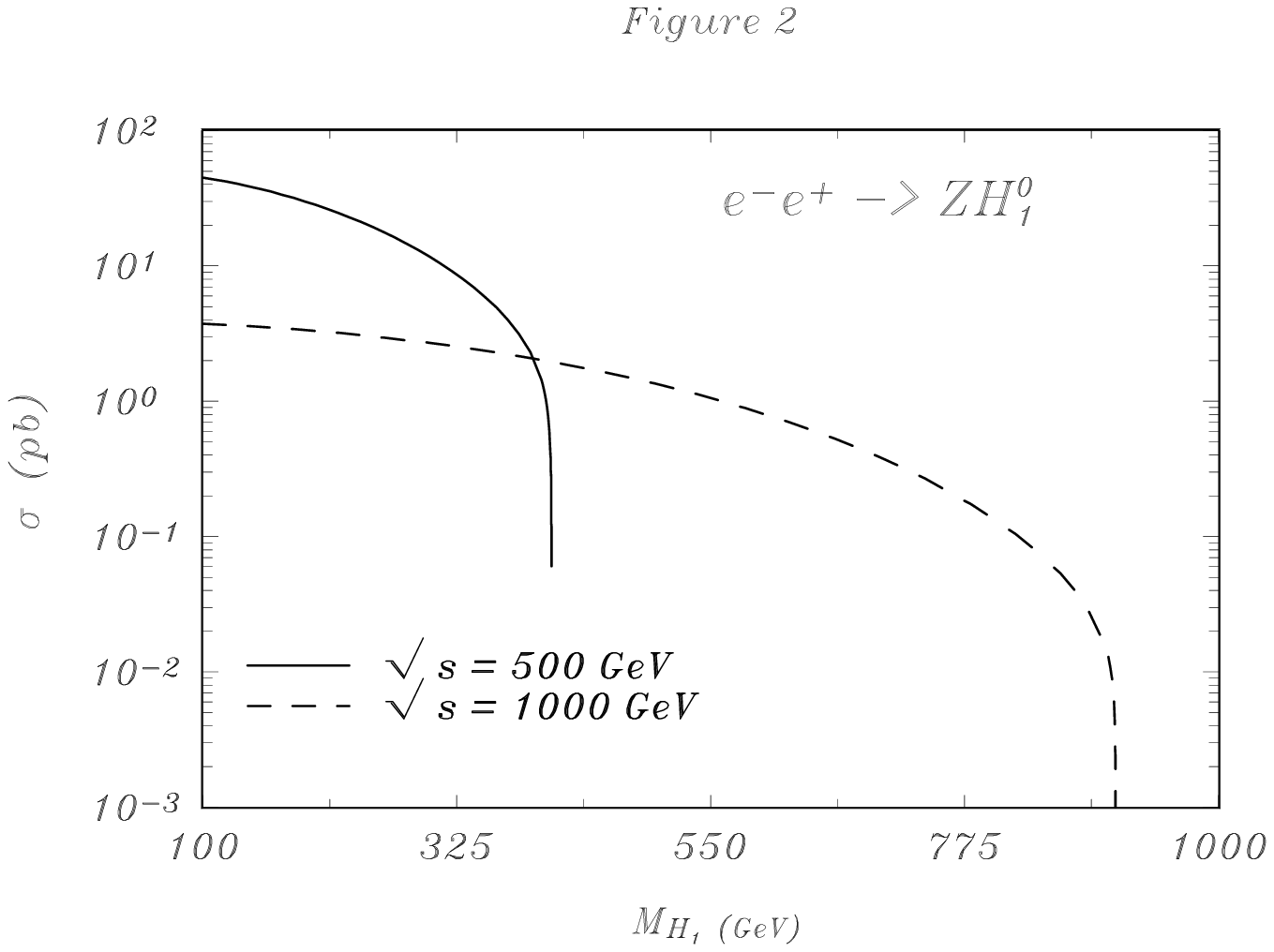}}
\caption{\label{fig2} \footnotesize Total cross section for the process $e^{-} e^{+} \rightarrow ZH_{1}^{0}$ as a 
function of $M_{H_{1}^{0}}$ for a $v_{\eta}= 240$ GeV at $\sqrt{s} = 500$ GeV (solid line) and $\sqrt{s} = 1000$ GeV (dashed line).}
\end{figure}
Considering that the expected integrated luminosity for both colliders will be of order of $6  \times 10^4$ pb$^{-1}$/yr and $2 \times 10^5$ pb$^{-1}$/yr, then the statistics we are expecting are the following. 
\begin{figure}
\scalebox{0.7}{\includegraphics*[0cm, 2.9cm][20cm,12cm]{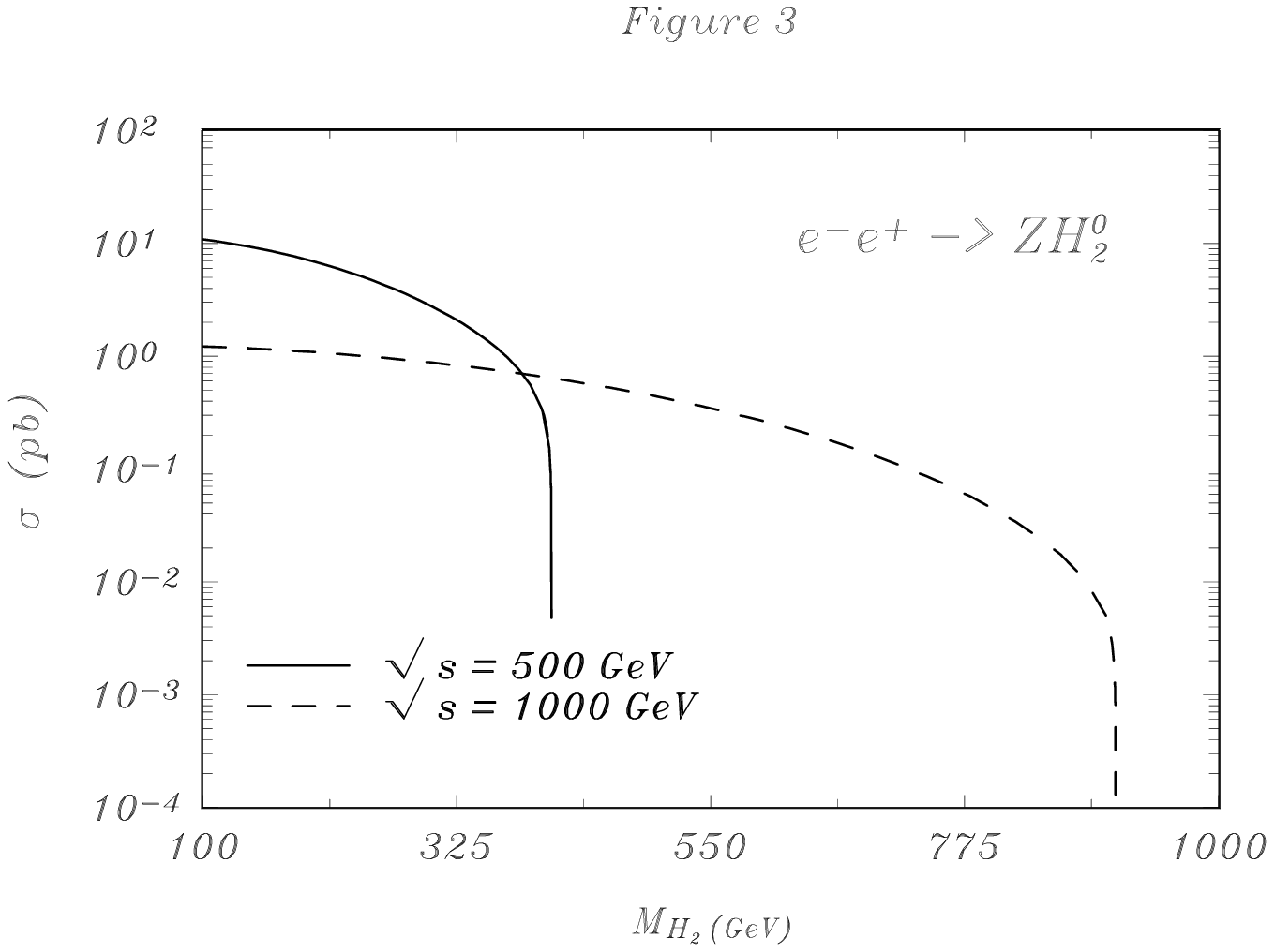}}
\caption{\label{fig3} \footnotesize Total cross section for the process $e^{-} e^{+} \rightarrow ZH_{2}^{0}$ as a 
function of $M_{H_{2}^{0}}$ for a $v_{\eta}= 140$ GeV at $\sqrt{s} = 500$ GeV (solid line) and $\sqrt{s}=1000$ GeV (dashed line).}
\end{figure}
The first  collider gives a total of $ \simeq 3.4 \times 10^4$ events per year for $v_{\eta}=140$ GeV, if we take the mass of the boson $M_{H^{0}_{1}}= 360$ GeV. 
\begin{figure}
\scalebox{0.7}{\includegraphics*[0cm, 2.9cm][20cm,12cm]{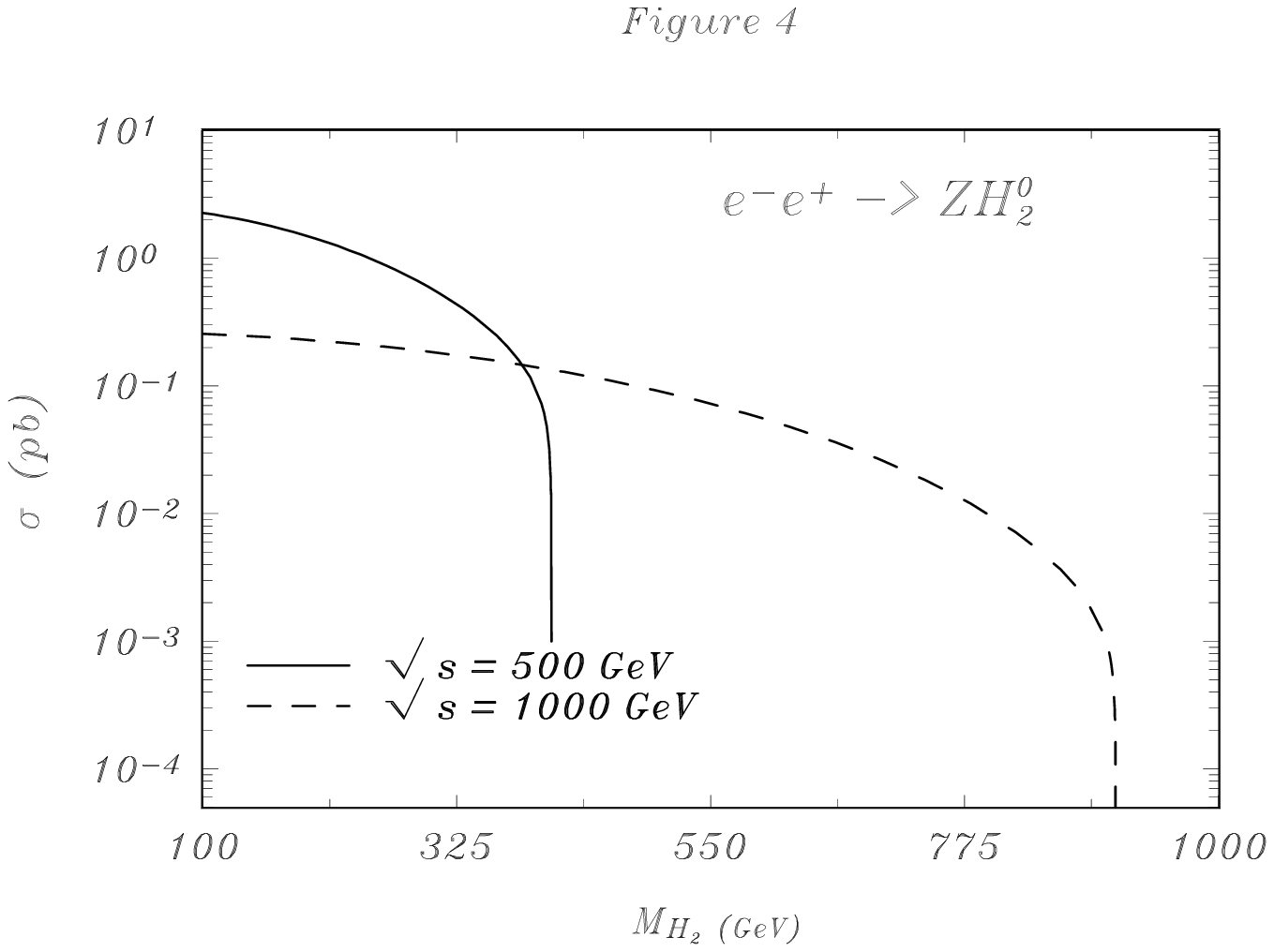}}
\caption{\label{fig4} \footnotesize Total cross section for the process $e^{-} e^{+} \rightarrow ZH_{2}^{0}$ as a 
function of $M_{H_{2}^{0}}$ for a $v_{\eta}=240$ GeV at $\sqrt{s} = 500$ GeV (solid line) and $\sqrt{s}=1000$ GeV (dashed line).}
\end{figure}
Considering that the signal for $H^{0}_{1}Z$ production will be $t \bar{t}$ and $q \bar{q}$ and taking into account that the branching ratios for both particles would be $B(H^{0}_{1} \to t \bar{t}) = 3.6 \%$ and $B(Z \to q \bar{q}) = 69.9 \%$, see Figs. 5 and 6, we would have approximately $855$ events per year. 
\begin{figure}
\scalebox{0.7}{\includegraphics*[0cm, 2.9cm][20cm,12cm]{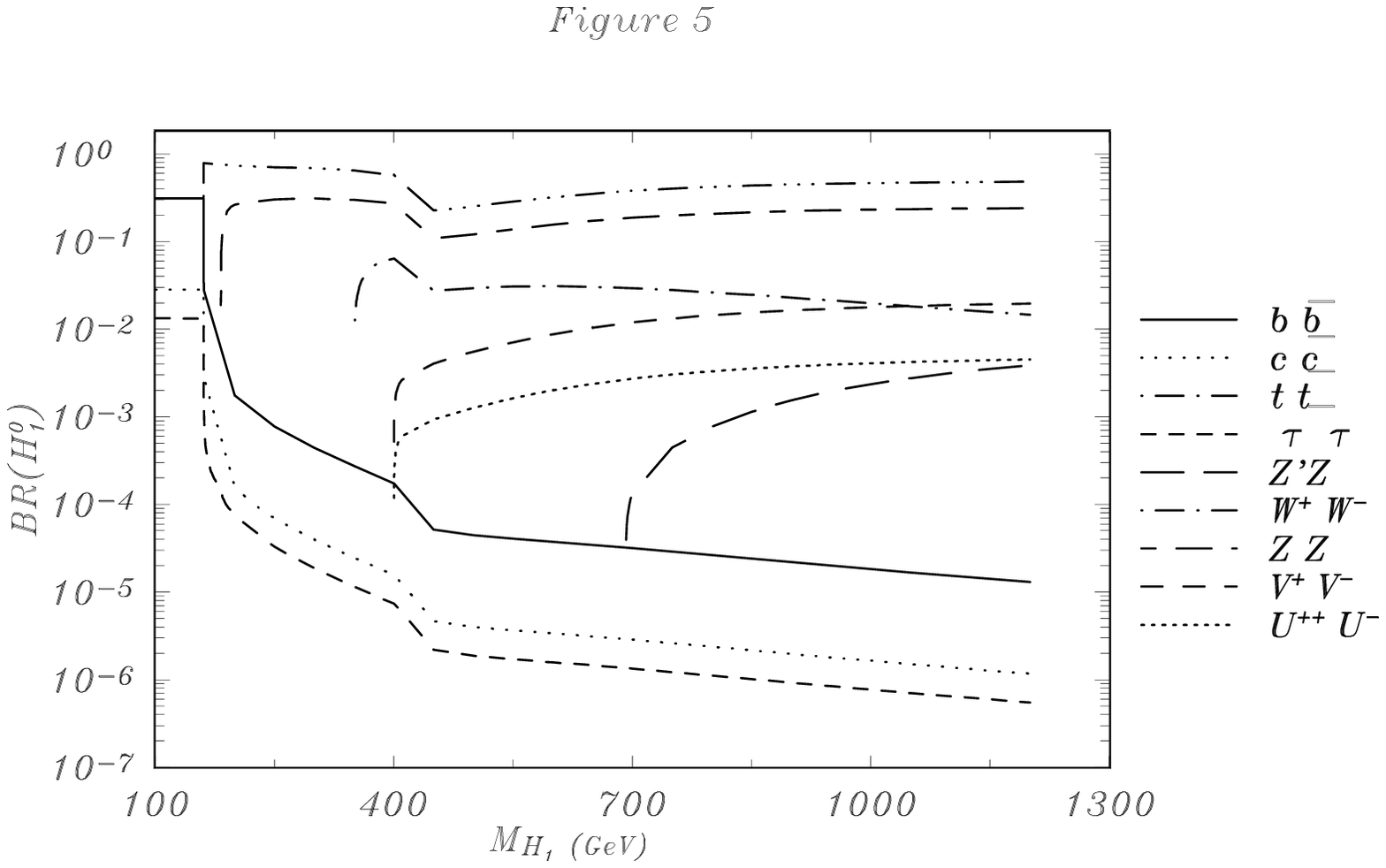}}
\caption{\label{fig5} \footnotesize Branching ratios for the Higgs decays as a functions of $M_{H_{1}^{0}}$ for $v_{\eta}= 140$ GeV.}
\end{figure}
Comparing this signal with the standard model background, like $e^{-} e^{+} \rightarrow W^{-} W^{+}, ZZ$, we note that this background can be easily distinguished and therefore eliminated by measuring the transverse mass of the two pairs of jets, see \cite{cazer}, but even so there is another small background, such as $e^{-} e^{+} \rightarrow WWZ$, but the cross section for this process is suppressed by at least $\alpha/\sin^{2} \theta_{W}$ relative to the process involving a double gauge boson, so using the COMPHEP \cite{pukhov}, the total cross section for this process will be equal to $4.23 \times 10^{-2}$ pb. 
\begin{figure}
\scalebox{0.7}{\includegraphics*[0cm, 2.9cm][20cm,12cm]{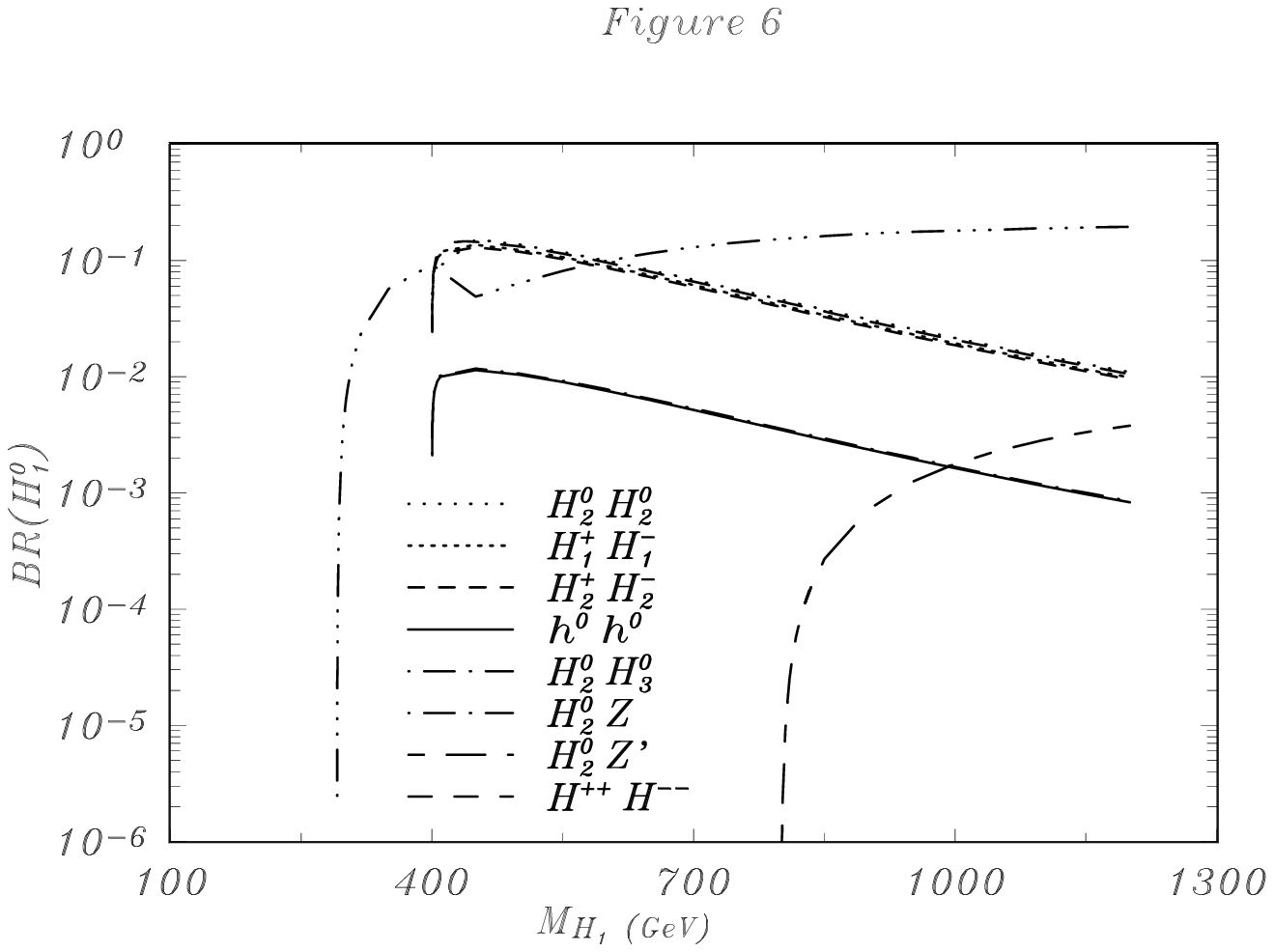}}
\caption{\label{fig6} \footnotesize Branching ratios for the Higgs decays as a functions of $M_{H_{1}^{0}}$ for $v_{\eta}= 140$ GeV.}
\end{figure}
The second collider (CLIC) gives a total of $\simeq 2.2 \times 10^{4}$ events per year if we take the same neutral Higgs mass, that is $M_{H^{0}_{1}}= 360$ GeV and considering the same branching ratios for the $H_{1}^{0}$ and the Z cited above, we would have nearly 553 events per year, for the signals and backgrounds see also \cite{cazer}.
\begin{figure}
\scalebox{0.7}{\includegraphics*[0cm, 2.9cm][20cm,12cm]{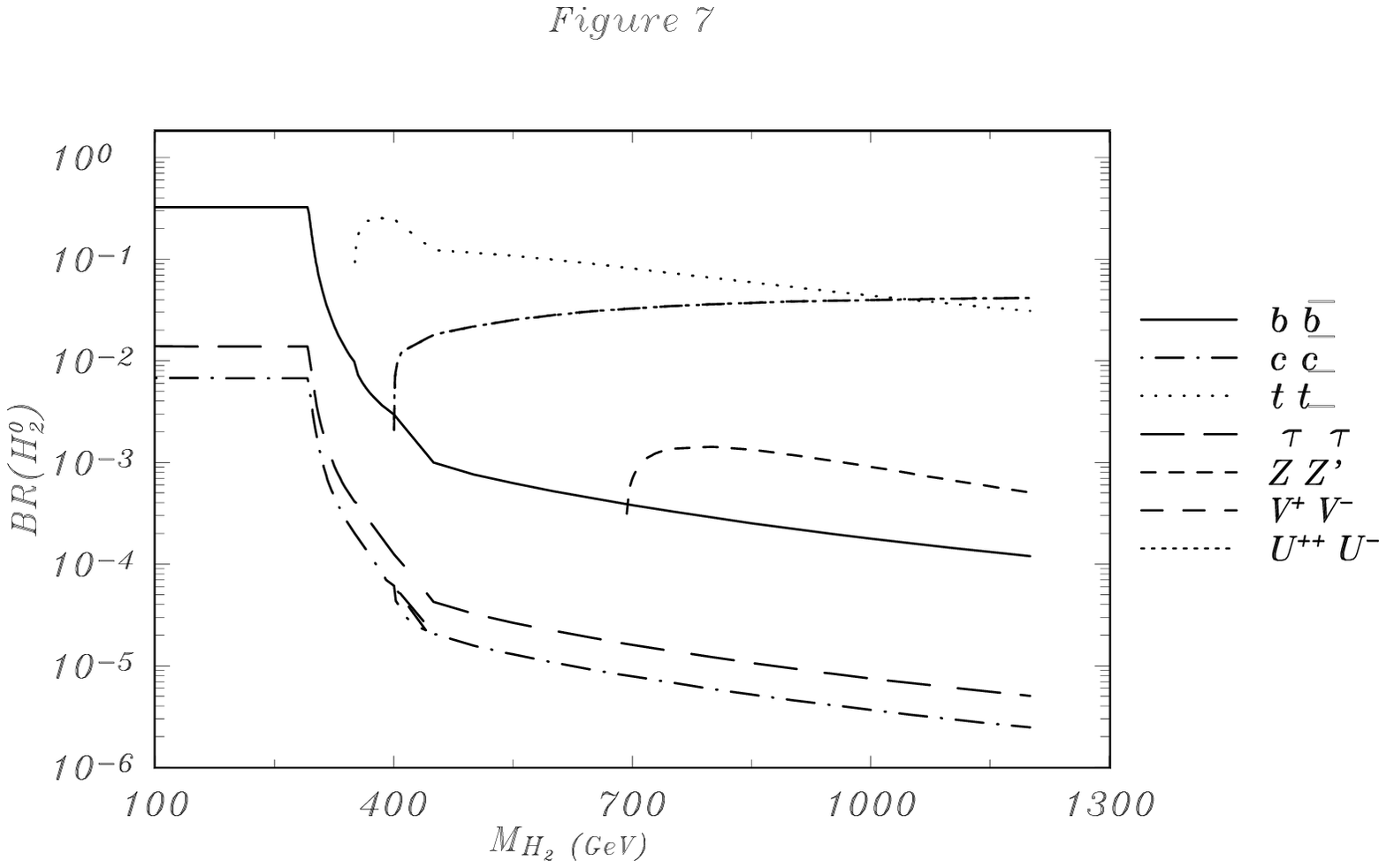}}
\caption{\label{fig7} \footnotesize Branching ratios for the Higgs decays as a functions of $M_{H_{2}^{0}}$ for $v_{\eta}= 140$ GeV.}
\end{figure}
In Fig. 2, we show the cross section for the production of the same particles as in Fig. 1, in the colliders NLC and CLIC for $v_{\eta}=240$ GeV and with $M_{H_{1}^{0}}=360$ GeV. We see from these results that we can expect for the first collider a total of $\simeq 1.05 \times 10^{7}$ events per year. 
\begin{figure}
\scalebox{0.7}{\includegraphics*[0cm, 2.9cm][20cm,12cm]{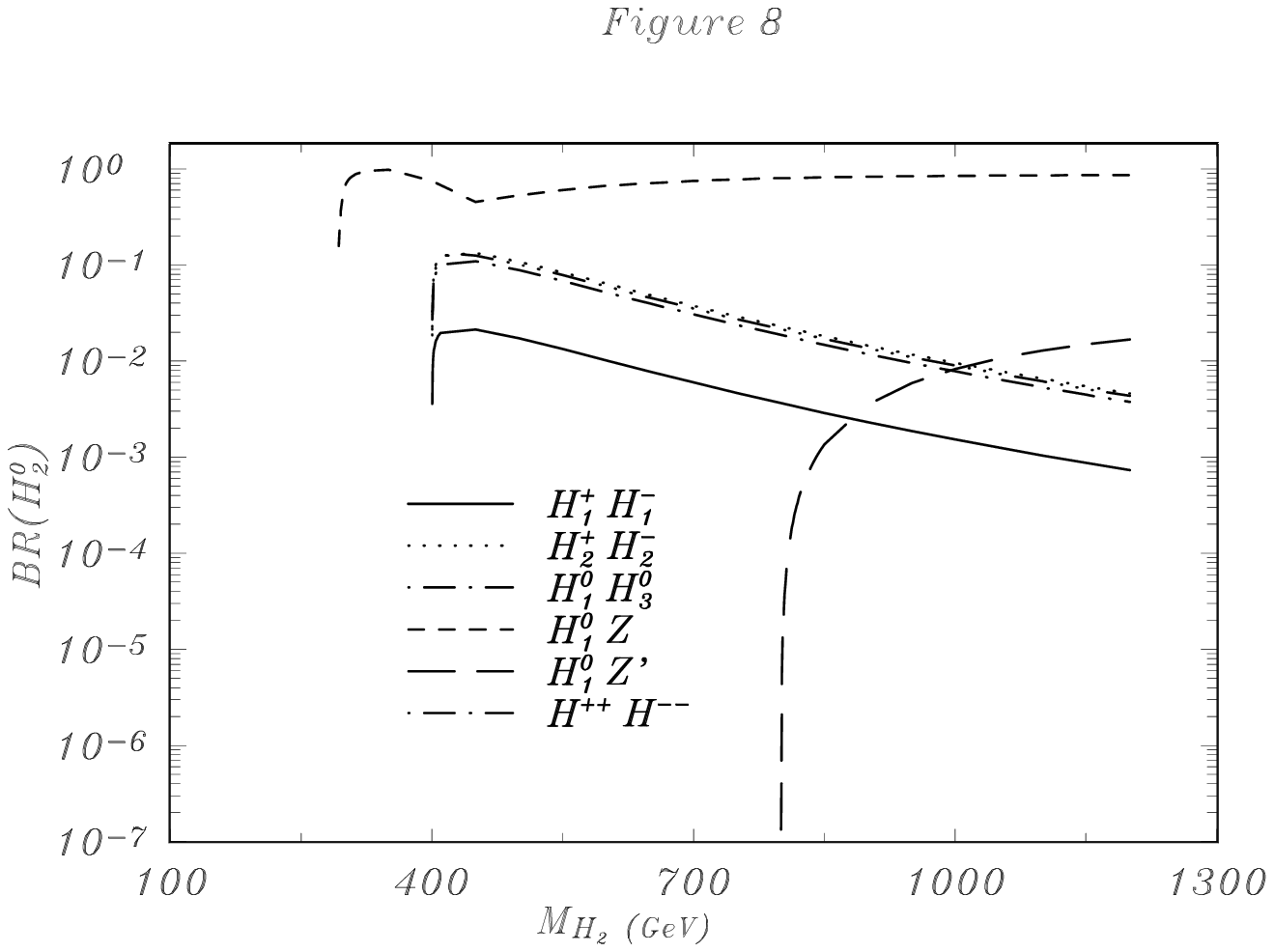}}
\caption{\label{fig8} \footnotesize Branching ratios for the Higgs decays as a functions of $M_{H_{2}^{0}}$ for $v_{\eta}=140$ GeV.}
\end{figure}
For the second collider, the CLIC, we expect a total of $2.9 \times 10^{5}$ events per year, that would be more than enough to establish the existence of the $H_{1}^{0}$. 
\begin{figure}
\scalebox{0.7}{\includegraphics*[0cm, 2.9cm][20cm,12cm]{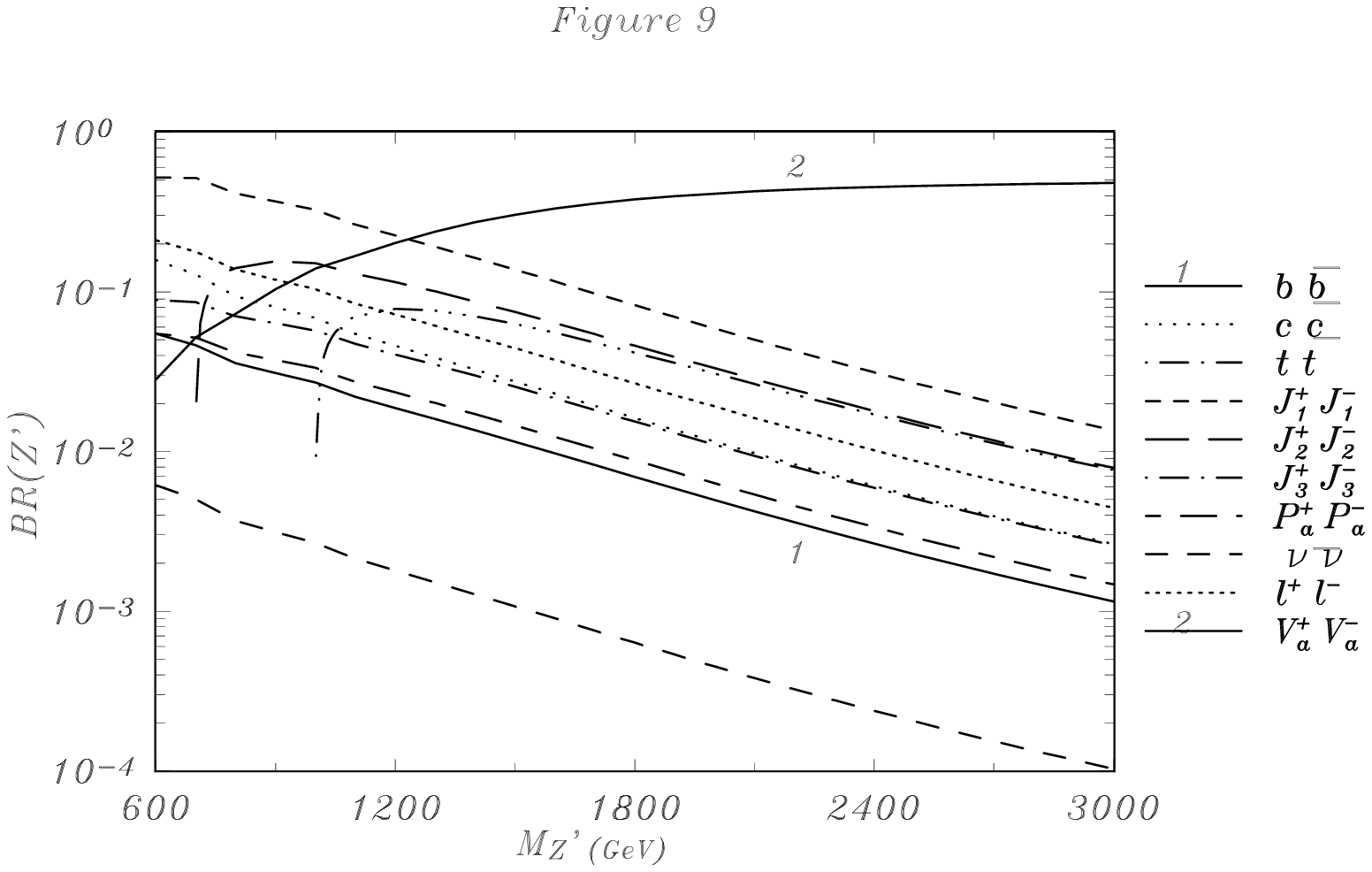}}
\caption{\label{fig9} \footnotesize Branching ratios for the Z$^{\prime}$ decays as a functions of $M_{Z^{\prime}}$ for $v_{\eta}=140$ GeV.}
\end{figure}
It is interesting to note the difference between the cross section for $v_{\eta}=140$ GeV and for $v_{\eta}=240$ GeV, this difference is due to the coupling constant, see (16).
\begin{figure}
\scalebox{0.7}{\includegraphics*[0cm, 2.9cm][20cm,12cm]{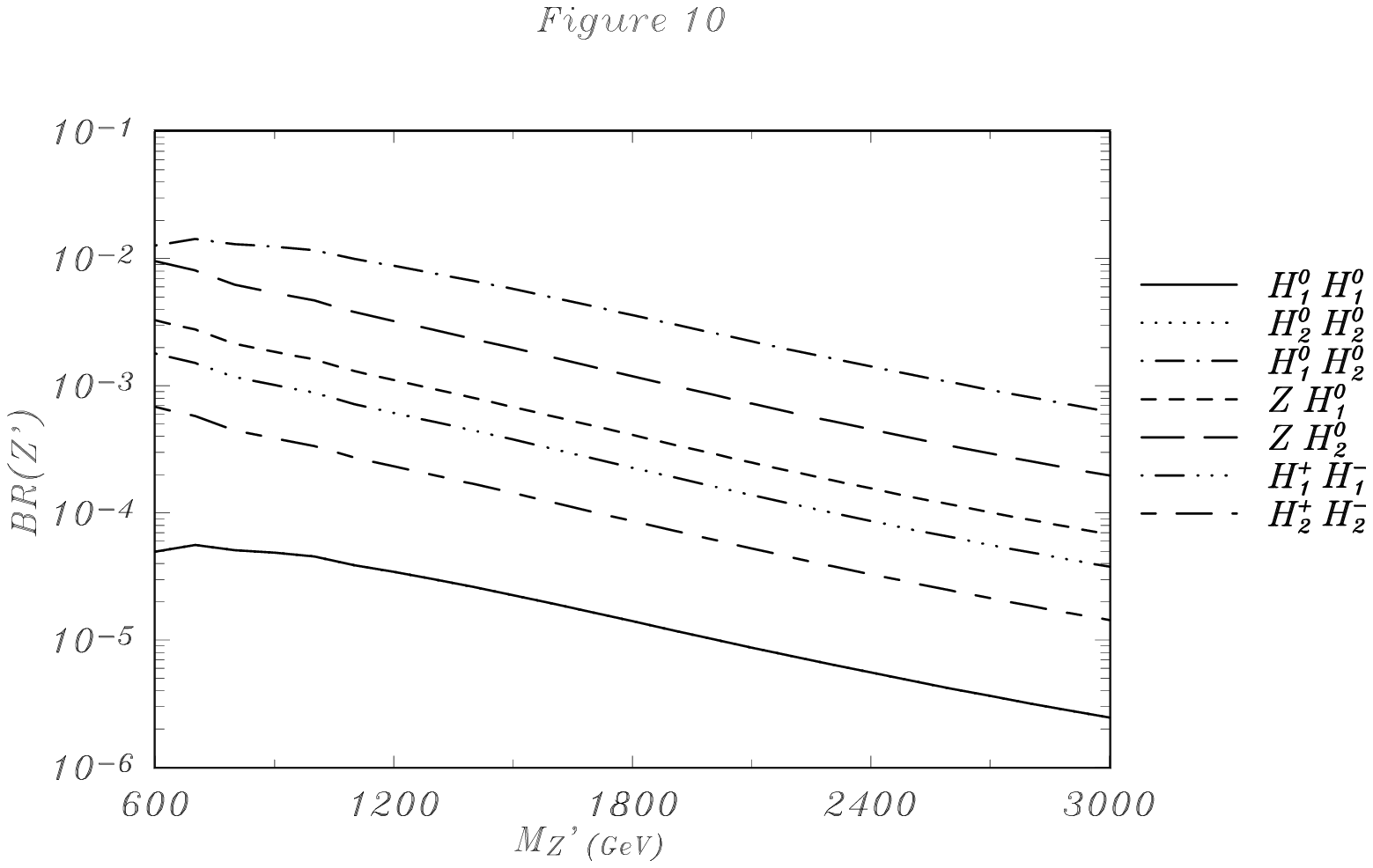}}
\caption{\label{fig10} \footnotesize Branching ratios for the $Z^\prime$ decays as a functions of $M_{Z^\prime}$ for $v_{\eta}= 140$ GeV.}
\end{figure}
In Fig. 3, it is shown the cross section for the production of $e^{-} e^{+} \rightarrow H_{2}^{0} Z$, for $v_{\eta} =140$ \ GeV with mass of $M_{H_{2}^{0}}= 360$ GeV. We see from these results that we can expect for the first collider a total of $\simeq 6.6 \times 10^{4}$ events per year to  produce $H_{2}^{0} Z$. Taking into account that the $H_{2}^{0}$ and Z will decay in $t \bar{t}$ and $\ell^{-} \ell^{+}$, see Figs. 7 and 8, and considering that the branching ratios for them are $B(H^{0}_{2} \rightarrow t \bar{t}) = 21.17 \%$ and $B(Z \rightarrow \ell^{-} \ell^{+}) = 3.36 \%$, then we will have a total of $\simeq 469$ events per year, however this events will be affected by backgrounds such as $e^{-} e^{+} \rightarrow q \bar{q}, WW, ZZ$ production, \cite{cazer}. For the second collider, the CLIC, we expect a total of $2.4 \times 10^{4}$ events per year for the mass of $H_{2}^{0}$ equal to $700$ GeV and $v_{\eta}=140$ GeV, considering now that the channel of decay will be $Z \rightarrow q \bar{q}$ and $H_{2} \rightarrow ZZ^\prime$ with $Z \rightarrow b \bar{b}$ and $Z^\prime \rightarrow e^{-} e^{+}$, which branching ratios are equal to $B( Z \rightarrow b \bar{b}) = 15,45 \%$ and $B( Z^\prime \rightarrow e^{-} e^{+}) = 5.9 \%$, see Figs. 9 and 10, we will have a total of $\simeq 153$ events per year, that is if we are looking for the signal \ $j \bar{j} \ b \bar{b} \ e \bar{e}$, we could discover the $H_{2}^{0}$ and $Z^\prime$, the backgrounds for this signal can be $WZZ$, $HZZ$, while the cross sections are so small $\propto 10^{-2}$ a detailed simulation of Monte Carlo must be done in all cases to extract the signal from the background.

Fig. 4 exhibits the total cross secction for the production of the same particles as in Fig. 3, in the colliders NLC and CLIC for $v_{\eta}=240$ GeV. We see from these results that we can expect for the first collider a total of $\simeq 1.5 \times 10^{4}$ events per year for $M_{H_{2}^{0}} =360$ GeV. This cross section is smaller compared with that of Fig. 3 by a factor of $0.227$. This difference between the cross section for $v_{\eta}=140$ and $v_{\eta}=240$ is due to the coupling constant (17). If we want to look for  a signal such as $e^{-} e^{+} \rightarrow H_{2}^{0} Z \rightarrow t \bar{t} \ell^{-} \ell^{+}$ we must multiply $469 \times 0.227$ that gives 106 events. We also have that the CLIC can produce a total of $5.3 \times 10^{3}$ for the mass of $M_{H_{2}}= 700$ GeV and for $v_{\eta}=240$ GeV, that is, this cross section is smaller by a factor of $0.22$ compared with the same process but for $v_{\eta}=140$ GeV, then the signal $e^{-} e^{+} \rightarrow H_{2}^{0} Z \rightarrow ZZ^\prime Z \rightarrow q \bar{q} (b \bar{b} \ e^{-} e^{+})$ would give a total of $\simeq 34$ events per year. So, we can conclude that the branching fraction measurements could tell us if the Higgs is standard or not.

We can also produce the Higgs bosons via the W-fusion, in which the Higgs bosons are formed in WW collisions and in association with neutrinos, that is

\[
e^{+} e^{-} \rightarrow H + \nu \bar{\nu} \ ,
\]
two mechanisms are responsible for this production, namely, Higgs-strahlung with Z decays to the three types of neutrinos and WW fusion \cite{cazer,petcov,altare,kramer,kniehl,boss}, being this last the dominant one for larger Higgs mass. Detailed analysis of this production will be given elsewhere \cite{cieto1}.

In summary, we have shown in this work that in the context of the 3-3-1 model the signatures for neutral Higgs bosons can be significant in both the NLC and in the CLIC colliders, however a detailed simulation of Monte Carlo must be done in all cases.

\acknowledgments

We would like to thank Prof. R. O. Ramos for a careful reading of the manuscript.


\end{document}